\documentclass[prl,aps,reprint,a4paper,superscriptaddress,longbibliography]{revtex4-2}
\usepackage{times}
\usepackage{amsfonts}
\usepackage{amsmath}
\usepackage{amssymb}
\usepackage{dsfont}
\usepackage{comment}
\usepackage{graphicx}
\usepackage{multirow}
\usepackage{float} 
\usepackage[colorlinks=true,linkcolor=blue,citecolor=blue,urlcolor=blue]{hyperref}

\begin{document}


\title{Simplest Kochen-Specker Set}


\author{Ad\'an~Cabello}
\email{adan@us.es}
\affiliation{Departamento de F\'{\i}sica Aplicada II, Universidad de Sevilla, E-41012 Sevilla,
Spain}
\affiliation{Instituto Carlos~I de F\'{\i}sica Te\'orica y Computacional, Universidad de
Sevilla, E-41012 Sevilla, Spain}


\begin{abstract}
Kochen-Specker (KS) sets are fundamental in physics. Every time nature produces bipartite correlations attaining the nonsignaling limit, or two parties always win a nonlocal game impossible to always win classically, is because the parties are measuring a KS set. The simplest quantum system in which all these phenomena occur is a pair of three-level systems. However, the simplest KS sets known in dimension three  are asymmetrical and require a large number of bases (the current minimum is 16, set by Peres and Penrose). Here we present a KS set that is much more symmetrical and easier to prove than any previous example. It sets a new record for minimum number of bases, 14, and enables us to refute Conjecture 2 in Phys. Rev. Lett. {\bf 134}, 010201 (2025), setting a new record for qutrit-qutrit perfect strategies with a minimum number of inputs: 5--9. We establish the fundamental nature of this set in quantum theory.
\end{abstract}


\maketitle


{\em Introduction---}A {\em Kochen-Specker (KS) set} \cite{Kochen:1967JMM} is a finite set of unit vectors $\mathcal{V}$ (or, equivalently, observables represented by rank-one projectors) in a Hilbert space ${\cal H}=\mathbb{C}^d$ of finite dimension $d \ge 3$, which does not admit an assignment $f: \mathcal{V} \rightarrow \{0,1\}$ such that $f(u) + f(v) \leq 1$ for $u, v \in \mathcal{V}$ orthogonal, and $\sum_{u \in b} f(u) = 1$ for every orthonormal basis $b \subseteq \mathcal{V}$.

KS sets are fundamental in physics for many reasons. They prove the impossibility of noncontextual hidden-variable theories \cite{Kochen:1967JMM}. They are at the basis of the experimental tests of nature's state-independent contextuality \cite{Cabello:2008PRL,Badziag:2009PRL,Kirchmair:2009NAT,Amselem:2009PRL,D'Ambrosio:2013PRX}, and of experiments producing nonlocality and contextuality simultaneously \cite{Cabello:2010PRL,Liu:2016PRL}. KS sets allow contextuality to be converted into nonlocality \cite{Cabello:2021PRL}. KS sets unique up to unitaries can be certified from their input-output statistics using {\em any} state of full rank \cite{Xu:2024PRL} and enable to Bell self-test supersinglets of any dimension \cite{Saha:2025XXX}. KS sets are {\em necessary} \cite{Cabello:2025PRL} for bipartite correlations attaining the nonsignaling limit \cite{Cabello:2001PRLb,ChenPRL2003,CinelliPRL2005,YangPRL2005}, and, via a result in \cite{Liu:2023XXX}, for bipartite perfect strategies (also known as ``quantum pseudotelepathy'') \cite{Xu:2022PRL}, and bipartite correlations with maximal nonlocal fraction (also known as ``fully nonlocal correlations'') \cite{Aolita:2012PRA}. In these last two roles, KS sets are crucial ingredients in some important results in quantum foundations and quantum computation, including the resolution of Tsirelson's problem \cite{Ji:2021CACM} and the quantum computational advantage in shallow circuits \cite{Bravyi:2018SCI}.

There are beautiful small KS sets in dimension four \cite{Cabello:1996PLA} and higher \cite{LisonekPRA2014}: they are symmetrical, the impossibility of a KS assignment is easy to prove, and each vector has nonzero components of equal magnitude. In contrast, none of the known small KS sets in dimension three, i.e., Sch\"utte-33 \cite{Bub:1996FP} (the number indicates the number of vectors), Peres-33 \cite{Peres:1991JPA}, Conway-Kochen-31 \cite{Peres:1993} (p.\ 114), and Penrose-33 \cite{Penrose:2000}, has any of these properties.


\begin{figure*}[hbt!]
 \centering
\includegraphics[width=0.69\linewidth]{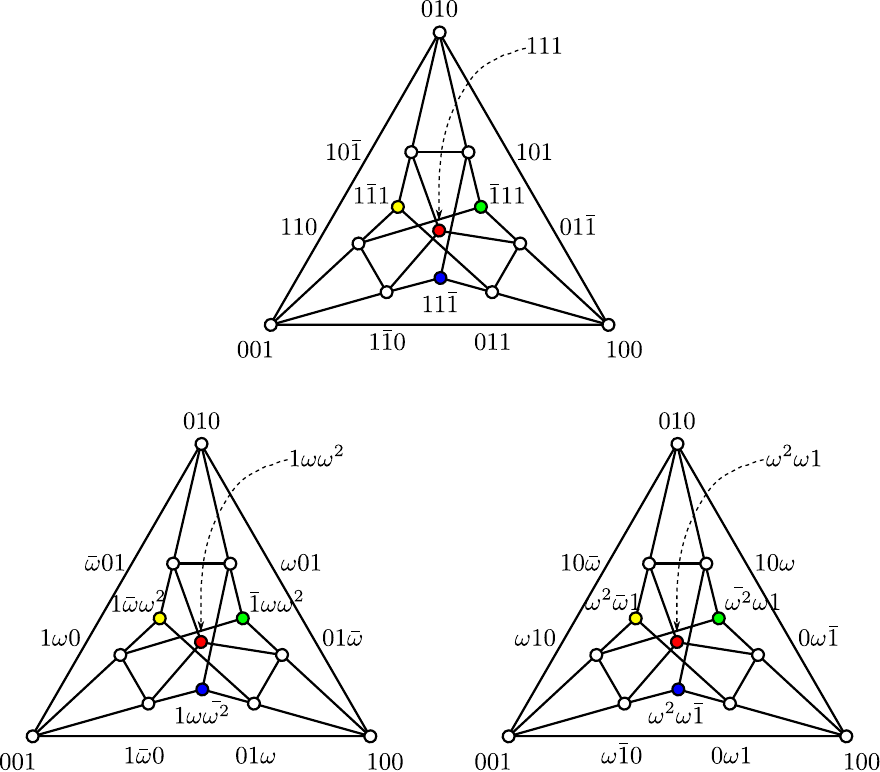}
 \caption{Relations of orthogonality between the 33 vectors of our KS set. Adjacent nodes represent orthogonal vectors. The three red (green, blue, yellow) nodes are mutually orthogonal.
 $1\omega\bar{\omega^2}$ represents $\tfrac{1}{\sqrt{3}}(1,\omega,-\omega^2)$, where $\omega = e^{2 \pi i /3}$.}
 \label{Fig1}
\end{figure*}


{\em Minimal-basis KS set---}Our set has 33 vectors. Their components and orthogonality relations are shown in Fig.~\ref{Fig1}. As conventional in quantum theory, vectors $|u\rangle$ and $|v\rangle$ are ``orthogonal'' when the inner product $\langle v |u\rangle=0$, where $\langle v |$ is the transpose {\em conjugate} of $|v\rangle$. The proof that they do not admit a KS assignment is as follows. The 13 vectors in the upper triangle in Fig.~\ref{Fig1} are the Yu-Oh set \cite{Yu:2012PRL} (see also \cite{Cabello:1996}, Fig.\ 2.9), which admits KS assignments only if {\em at most one} of the four nonwhite nodes is assigned $1$. 
The fact that no KS assignment is possible for all vectors in Fig.~\ref{Fig1} follows from the observation that any assignment would require exactly one of the red (or green or blue or yellow) nodes to be assigned $1$ because the three red (blue, green, yellow) nodes are mutually orthogonal.
However, that is impossible because there is only room for {\em three} $1$'s (one per Yu-Oh set), while there are {\em four} colors (red, green, blue, and yellow). 

As frequently pointed out (see, {\em e.g.}, \cite{LisonekPRA2014} and references therein), the true measure of physical simplicity of a KS set is its number of bases. This has been corroborated by the recent discovery that, for the simplest KS sets,
what determines the sum of Alice and Bob's inputs necessary to attain a perfect quantum strategy is the number of {\em complete} orthogonal bases 
\cite{Cabello:2025PRL,Trandafir:2025PRAb} [{\em e.g.}, $\{(0,0,1), (0,1,0), (1,0,0)\}$ is a complete orthogonal basis, while the vector needed to complete an orthogonal basis containing $\tfrac{1}{\sqrt{2}}(0,1,-1)$ and $\tfrac{1}{\sqrt{3}}(1,1,1)$ is not in the KS set]. The original KS set \cite{Kochen:1967JMM} had $43$ complete bases \cite{Budroni:2022RMP} (Fig.\ 1). Peres set the current record, $16$ complete bases, in 1991 \cite{Peres:1991JPA}. Penrose equaled the record in 2000 with a different construction \cite{Penrose:2000}. Conway-Kochen-31 (Ref.\ \cite{Peres:1993}, p.\ 114) has $17$ complete bases and Sch\"utte-33 \cite{Bub:1996FP} has $20$ complete bases. The set introduced in this Letter sets a new record: $14$. They correspond to the large triangle in Fig.~\ref{Fig1} (which is repeated three times), the $9$ small triangles, and the four colored bases. 

It should be noted that some authors use a different definition of KS set that requires additional vectors and bases \cite{Trandafir:2025XXX}.


{\em Qutrit-qutrit perfect strategy with minimal settings---}The KS introduced here allows us to refute Conjecture 2 in \cite{Cabello:2025PRL}, which stated that the qutrit-qutrit perfect strategy with minimum $|X| \times |Y|$, where $|X|$ is Alice's input cardinality and $|Y|$ is Bob's, has $|X|=7$ and $|Y|=9$, and corresponds to distributing the 16 bases of Peres-33 (or Penrose-33) in a particular way. 

Using the new KS set, we can define a bipartite nonlocal game with $|X|=5$ and $|Y|=9$ and construct a qutrit-qutrit perfect strategy for it as follows. With probability $\pi(x,y)$, Alice is given a basis $x \in X$, where $X$ has the following five elements (for simplicity, hereafter, vectors may be un-normalized):
\begin{subequations}
\begin{align}
   x=0:\;\;& \{(0,0,1),(0,1,0),(1,0,0)\}, \label{type1} \\
   x=1:\;\;& \{(1,\omega,\omega^2),(1,1,1),(\omega^2,\omega,1)\}, \label{type2a} \\
   x=2:\;\;& \{(1,\omega,-\omega^2),(1,1,-1),(\omega^2,\omega,-1)\}, \\
   x=3:\;\;& \{(1,-\omega,\omega^2),(1,-1,1),(\omega^2,\omega,1)\}, \\
   x=4:\;\;& \{(-1,\omega,\omega^2),(-1,1,1),(-\omega^2,\omega,1)\} \label{type2b},
\end{align}
\end{subequations}
with $\omega = e^{2 \pi i /3}$, and Bob is given a basis $y \in Y$, where $Y$ has the following nine elements:
\begin{subequations}
\begin{align}
    y=0:\;\;& \{(0,0,1),(1,1,0),(1,-1,0)\}, \label{type3a} \\
    y=1:\;\;& \{(0,0,1),(1,\omega,0),(1,-\omega,0)\}, \\
    y=2:\;\;& \{(0,0,1),(\omega,1,0),(\omega,-1,0)\}, \\
    y=3:\;\;& \{(0,1,0),(1,0,1),(1,0,-1)\}, \\
    y=4:\;\;& \{(0,1,0),(1,0,\omega),(1,0,-\omega)\}, \\
    y=5:\;\;& \{(0,1,0),(\omega,0,1),(\omega,0,-1)\}, \\
    y=6:\;\;& \{(1,0,0),(0,1,1),(0,1,-1)\}, \\
    y=7:\;\;& \{(1,0,0),(0,1,\omega),(0,1,-\omega)\}, \\
    y=8:\;\;& \{(1,0,0),(0,\omega,1),(0,\omega,-1)\}. \label{type3b}
\end{align}
\end{subequations}
Then, the referee asks each player to choose a vector from the basis they were given.
Alice and Bob win if, and only if, their output vectors are not orthogonal. To understand the connection between this game and the impossibility of a KS assignment, observe that there are $5 \times 9$ possible combinations of Alice's and Bob's bases (i.e., there are $45$ contexts). These combinations fall into two distinct types.

The first type comprises nine combinations where Alice's and Bob's bases share a common vector. Then, Alice and Bob win if, and only if, they make the same KS assignment to the common vector. This can occur in two ways: both of them output the common vector (meaning both assign $f=1$ to the common vector and $f=0$ to the other two vectors in their respective bases), or both output a vector different from the common one (meaning both assign $f=0$ to the common vector). Consequently, there are five combinations of outputs that result in a win and four combinations that result in a loss.

The second type comprises $36$ combinations of bases where one vector in Alice's basis is orthogonal to one vector in Bob's basis. Here, Alice and Bob win except when they output the two orthogonal vectors. Consequently, there are eight combinations of outputs that result in a win and one combination that results in a loss.

Assuming that all inputs are equally distributed, the winning probability is given by the sum of the probabilities of the $9 \times 5 + 36 \times 8 = 333$ winning combinations of outputs, divided by the number of possible combinations of bases. The classical winning probability cannot be one, because the set of vectors is a KS set. In fact, the maximal classical winning probability is given by the independence number of the corresponding $333$-vertex orthogonality graph divided by $45$ \cite{CSWPRL2014}. That is,
\begin{equation}
    W_C = \frac{44}{45} \approx 0.9778.
\end{equation}
However, suppose that Alice and Bob share the maximally entangled qutrit-qutrit state 
\begin{equation}
    |\psi\rangle = \frac{1}{\sqrt{3}} \sum_{j=0}^2 |j\rangle \otimes |j\rangle,
\end{equation}
and each player measures their qutrit in the basis provided by the referee and outputs the vector corresponding to their result. This strategy always produces output vectors that are not orthogonal. Therefore, the quantum winning probability is
\begin{equation}
    W_Q = 1.
\end{equation}

Moreover, by using the the method in \cite{Trandafir:2025PRAb}, it can be proven that no other way of distributing the elements of the KS set produces a perfect quantum strategy with smaller $|X| \times |Y|$. Remarkably, the symmetries in the KS set's orthogonality graph determine the distribution of bases for achieving minimality. The connectivity of the graph classifies the complete bases into three types. (i) Eq.~\eqref{type1}, (ii) Eqs.~\eqref{type2a}--\eqref{type2b}, and (iii) Eqs.~\eqref{type3a}--\eqref{type3b}. The perfect qutrit-qutrit strategy with minimal $|X| \times |Y|$ is realized when Alice measures types i and ii, and Bob measures type iii.


{\em Why is this set fundamental?---}We argue that the KS set presented here is the simplest possible for a qutrit and a fundamental object in quantum theory. Our claims are supported by the following evidence:

(a) It contains fewer complete bases than any prior known set.

(b) It is significantly more symmetrical, as quantified in Table~\ref{table1}.

(c) It yields a perfect qutrit-qutrit strategy with a lower input cardinality than any known set.

(d) It is unique up to unitary transformations, whereas neither Peres-33 nor Penrose-33 are \cite{Penrose:2000,Gould:2010FPH,bengtsson2012gleason,Blanchfield:2014,Xu:2022PRL}. This uniqueness can be proven using the method described in \cite{Xu:2022PRL}. For an explicit proof, see \cite{Trandafir:2026PRA}.

(e) While all KS sets demonstrate state-independent contextuality (SI-C), the converse is not true; not all vector sets exhibiting SI-C are KS sets. The Yu-Oh set \cite{Yu:2012PRL} (the 13 vectors at the top of Fig.~\ref{Fig1}) was the first such example. On the other hand, every known small KS set contains the Yu-Oh set. This is reasonable, as the Yu-Oh set has been proven to be the minimal set of vectors (in any dimension) that allows for SI-C \cite{Cabello:2016JPA}. It is therefore natural to conjecture that the simplest qutrit KS set must contain the Yu-Oh set.
An analysis of the Yu-Oh set's orthogonality graph reveals that its vertices fall into three distinct types based on their connectivity. Therefore, any KS set containing the Yu-Oh set must possess at least three vertex types. All previously known KS sets have more than three types (see Table~\ref{table1}). The KS set introduced here is the first with exactly three types: (I) he three vertices corresponding to the vectors in Eq.~\eqref{type1}, (II) the $12$ vertices in Eqs.~\eqref{type2a}--\eqref{type2b} that are not type I, and (III) the 18 vertices in Eqs.~\eqref{type3a}--\eqref{type3b} that are not type I.

(f) As observed by Ingemar Bengtsson \cite{Ingemarpriv}, the KS set is obtained by applying the generators of the Weyl-Heisenberg group---whose algebra is defined by the canonical commutation relations \cite{Weyl:1931}---to the minimal SI-C set. Specifically, in $d = 3$, the Weyl-Heisenberg group is generated by
\begin{equation}
X = \begin{pmatrix}
0 & 0 & 1 \\
1 & 0 & 0 \\
0 & 1 & 0
\end{pmatrix},\;\;\;\;\;\;
Z = \begin{pmatrix}
1 & 0 & 0 \\
0 & \omega & 0 \\
0 & 0 & \omega^2
\end{pmatrix}.
\end{equation}
If we act with $X$ on the simplest SI-C set, we stay within the simplest SI-C set. If we act with $Z$ twice, we obtain the simplest KS set.
The fact that this combination of two fundamental structures in quantum theory (the Weyl-Heisenberg group, which is at the center of quantum theory [\cite{Schwinger:2001}, p.\ 74], and the minimal SI-C set) yields a KS set, strongly reinforces our view that this set is a fundamental object.

(g) It contains two unitarily inequivalent symmetric, informationally complete, positive operator-valued measures (SIC-POVMs): If we act with both $X$ and $Z$ on $\tfrac{1}{\sqrt{2}}(1,1,0)$, we obtain a SIC-POVM. If we act with both $X$ and $Z$ on $\tfrac{1}{\sqrt{2}}(1,-1,0)$, we obtain another SIC-POVM \cite{Ingemarpriv}. 

\begin{figure}[hbt!]
 \centering
\includegraphics[width=0.86\linewidth]{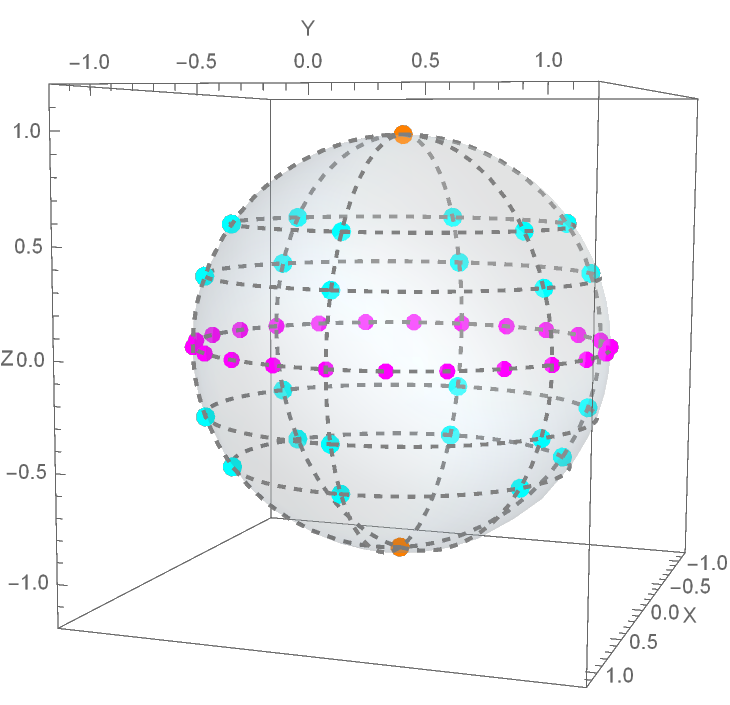}
 \caption{Majorana representation of the KS set. Each of the $33$ vectors corresponds to a unique pair of points, but some points are shared among different vectors.}
 \label{Fig2}
\end{figure}

{\em Geometry---}Following Penrose \cite{Penrose:2000}, we can visualize the $33$ complex unit vectors of the KS sets using the Majorana representation \cite{Majorana:1932NC}. In the Majorana representation every pure state of spin-$1$ (i.e., complex unit vector in dimension three) is uniquely represented by {\em two} points (possibly coinciding) on the unit sphere. The meaning of these points is the following. A spin-$1$ state can be seen as a symmetrized combination of two spin-$1/2$'s. The two points on the Bloch sphere represent the directions of these two spin-$1/2$'s. For example, if the two points coincide then the spin-$1$ state is a coherent state (maximally polarized along some direction). If the two points are antipodal, then the spin-$1$ state has no net dipole moment. For example, $(1,0,0)$ is represented by two points in the north pole, $(0,1,0)$ is represented by one point in the north pole and one point in the south pole, and $(0,0,1)$ is represented by two points in the south pole.
Details of the representation can be found in \cite{Penrose:2000} and Sec.\ 4.4 of \cite{BengtssonZyczkowski:2017}. Figure~\ref{Fig2} shows all points in the Majorana representation of our $14$-basis $33$-vector KS set.


{\em Conclusions---}For almost 35 years, Peres-33 \cite{Peres:1991JPA} has been the ``standard'' KS set: the one with the fewest bases, the one with the highest degree of symmetry, the one that determines the simplest scenario in which nature reaches the nonsignaling limit, produces perfect nonlocal strategies, and generates fully nonlocal correlations using a bipartite system of minimal dimension. Peres-33 has been the default option for a KS set. It is the one that is on the cover of modern books \cite{Halvorson:2011}. It is the one used to explain the use of KS sets in nonlocal games \cite{Mancinska:2007} to the point that even Conway and Kochen used Peres-33 (instead of Conway-Kochen-31) to prove both the free-will theorem \cite{CK06} and the strong free-will theorem \cite{CK09}.

In this Letter, we have introduced a KS set that substantially improves Peres-33 in all aspects. Table~\ref{table1} enables us to appreciate the improvement. In a nutshell, our result shows that nature does it better than we thought. A KS state-independent contradiction with noncontextual hidden-variable theories only requires $14$ bases. These bases define a much more elegant (symmetrical) qutrit KS set than any previous one. Nature only needs Alice to choose between five settings and Bob to choose between nine settings to produce, with pairs of systems of minimal dimension (qutrits), correlations attaining the nonsignaling limit, allowing them to always win a game that is impossible to always win with classical methods, and showing full nonlocality (i.e., nonlocal content $1$, where nonlocal content is defined in Ref.\ \cite{Elitzur:1992PLA}). 


\begin{table*}[]
\centering
\begin{tabular}{lcccc}
\hline \hline
& Bases & Types of vertices & Degree of symmetry & Minimal bipartite perfect strategy \\
\hline
Sch\"utte-33 \cite{Bub:1996FP} & $20$ & $9$ & $8$ & $8-9$ \\
Conway-Kochen-31 \cite{Peres:1993} & $17$ & $10$ & $4$ & $8-9$ \\
Peres-33 \cite{Peres:1991JPA} & $16$ & $4$ & $48$ & $7-9$\\
Penrose-33 \cite{Penrose:2000} & $16$ & $4$ & $48$ & $7-9$\\
This Letter & $14$ & $3$ & $144$ & $5-9$ \\
\hline\hline
\end{tabular}
\caption{Simplest KS sets in dimension three. Bases gives the number of complete bases. The number of types of vertices is obtained as follows. The automorphisms of the orthogonality graph (i.e., the permutations of vertices that preserve adjacency) induce a partition of the vertices into orbits. Two vertices belong to the same orbit if, and only if, there exists an automorphism that takes one
to the other. Each of the orbits contains vertices that are structurally equivalent. The number of types of vertices is the number of orbits,
The degree of symmetry is quantified by the size of the automorphism group of the orthogonality graph. The minimal bipartite perfect strategy is the perfect strategy that minimizes $|X| \times |Y|$, where $|X|-|Y|$ indicates Alice's and Bob's number of settings, respectively. The minimal perfect strategies have been obtained using the method in \cite{Trandafir:2025PRAb}.}
\label{table1}
\end{table*}


The fact that it has taken us 35 years to make progress and that the remaining margin for improvement is shrinking thanks to new methods for exploring bipartite perfect quantum strategies \cite{Liu:2023XXX,Cabello:2025PRL,Trandafir:2025PRAb} and finding lower bounds on the size of the smallest KS set \cite{Kirchweger:2023,Li:2024}, together with the reasons provided before, may indicate that we are already at the true natural limit. We are actively working to prove this conjecture.


{\em Acknowledgments---}I thank Ingemar Bengtsson, Curtis Bright, Vijay Ganesh, Zhengyu Li, Karl Svozil, and Stefan Trandafir for helpful discussions. I also thank Stefan Trandafir for computing the sizes of the automorphism groups appearing in Table~\ref{table1} using SageMath. This work was supported by EU-funded project \href{10.3030/101070558}{FoQaCiA}.


{\em Data availability---}The data that support the findings of this article are openly available \cite{data}.



%


\end{document}